\documentclass{ws-ijmpd}
\usepackage{amssymb}

\begin{document}

\markboth{Xiaoyu Lai and Renxin Xu} {Quark-cluster Stars: the
structure}

\catchline{}{}{}{}{}
%
%%%%%%%%%%%%%%%%%%%%%%%%%%%%%%%%%%%%%%%%%%%%%%%%%%%%%%%%%%%%%%%%%%%%

%\title{PULSARS: QUARK (CLUSTER) STARS}
\title{Quark-cluster Stars: the structure}

\author{XIAOYU LAI \and RENXIN XU}

\address{School of Physics and State Key Laboratory of Nuclear Physics and Technology,\\
Peking University, Beijing, 100871, China; %
\{laixy, r.x.xu\}@pku.edu.cn}

\maketitle

%\begin{history}
%\received{Day Month Year}
%\revised{Day Month Year}
%\comby{Managing Editor}
%\end{history}

\begin{abstract}
The nature of pulsar-like compact stars is still in controversy
although the first pulsar was found more than 40 years ago.
Generally speaking, conventional neutron stars and non-mainstream
quark stars are two types of models to describe the inner structure
of pulsars, with the former composed mainly of hadrons and the
latter of a peculiar kind of matter whose state equation should be
understood in the level of quarks rather than hadrons.
To construct a more realistic model from both theoretical and
observational points of view, we conjecture that pulsars could be
``quark-cluster stars'' which are composed of quark-clusters with
almost equal numbers of up, down and strange quarks.
Clustering quark matter could be the result of strong coupling
between quarks inside realistic compact stars. The lightest quark
clusters could be of $H$-dibaryons, while quark clusters could also
be heavier with more quarks. Being essentially related to the
non-perturbative quantum-chromo dynamics (QCD), the state of
supra-nuclear condensed matter is really difficult to obtain
strictly by only theoretical QCD-calculations, and we expect,
nevertheless, that astrophysical observations could help us to have
a final solution.
\end{abstract}


\section{Introduction}

The subjects which are attractive and of great importance are
usually those ones that are beyond our comprehension, and the nature
of pulsar-like compact stars is of such a kind.
Our understanding towards pulsars is still developing, thanks to the
developments both theoretical and observational, but the real state
of matter is still uncertain. Neutron stars and quark stars, as two
types of models for the nature of pulsar, have been debated for a
very long time, but which model is more realistic remains to be
seen.

The research history of extremely dense matter goes back to the very
early time of stellar compact objects.
Astronomers were not be able to understand the compactness of white
dwarfs until the British physicist Ralph Howard Fowler (1889$-$1944)
recognized the quantum pressure of degenerate electrons there, who
for the first time discussed also dense matter at his maximum
possible density in the same seminal paper (1926)~\cite{Fowler1926}
as ``{\em The density of such `energetic' matter is then only
limited a priori by the `sizes' of electrons and atomic nuclei. The
`volumes' of these are perhaps $10^{-14}$ times of the volume of the
corresponding atoms, so that densities up to $10^{14}$ times that of
terrestrial materials may not be impossible}'' after Ernest
Rutherford constructed the nucleus model of atom in 1911.
What if the matter density becomes so high?

The year 1932 was special.
(1) Neutron (or ``neutral doublet'' in Rutherford's words) was
experimentally discoveries by James Chadwick although it had been
speculated to exist for a long time and for a variety of reasons.
(2) Landau~\cite{Landau1932} conjectured a condensed core with
nuclear matter densities inside a star where protons and electrons
combined tightly forming the ``neutronic'' state in order to explain
the origin of stellar energy.
In addition to advanced and detailed calculations, authors are
modeling normal neutron stars generally along Landau's line although
Landau did make two mistakes~\cite{xu11} 80 years ago because of the
historical limitations.

The inner structure of quark stars which are totally composed of
quark matter (with $u$, $d$ and $s$ quarks) was first calculated by
Itoh~\cite{Itoh1970} in 1970,  because it was realized previously
that there could be deconfined quarks inside neutron stars. From
then on, neutron stars are defined as such a kind of compact objects
that mainly composed of neutrons, with hyperons or even quark matter
in their innermost cores, and quark stars as a kind of compact
objects composed of pure (strange) quark matter. It is worth
mentioning that, quark stars are characterized by soft equations of
state, because the asymptotic freedom of QCD tells us that as energy
scale goes higher, the interaction between quarks becomes weaker.
The recent discovery~\cite{2Msun} of a $\sim 2M_\odot$ neutron star
seems to be evidence against quark stars unless the coupling between
deconfined quarks is still very strong.~\cite{Alford2007}
Anyway, a working model with the coupling parameter as high as
$\alpha_s\gtrsim 0.6-0.7$ could be possible in
principle,~\cite{massive} but such a strong interaction may also
favour a kind of condensation in {\rm position} space rather than in
momentum space as was already noted in 2003.~\cite{Xu03}

Neutron stars and quark stars respectively correspond to two
distinct regions in the QCD phase diagram, the hadron phase in the
low density region and the quark-gluon plasma phase in the high
density region.
In other words, inside neutron stars the highly non-perturbative
strong interaction makes quarks grouped into neutrons, whereas
inside quark stars the perturbative strong interaction makes quarks
to be almost free if the coupling is weak. At a few nuclear matter
densities and extremely low temperature, the quark degrees of
freedom should be significant, and there is possible observational
evidence that pulsars could be quark stars (see reviews, e.g.
Ref~\refcite{weber05,xu08a,xu08b}). However, in cold quark matter at
realistic baryon densities of compact stars ($\rho\sim 2-10\rho_0$),
the energy scale is far from the region where the asymptotic freedom
approximation could apply.
The strong coupling between quarks even exists in the hot
quark-gluon plasma~\cite{Shuryak2009}, then it is reasonable to
infer that quarks could be coupled strongly also in the interior of
quark stars, which could make quarks to condensate in position space
to form quark clusters. The quark matter inside compact stars could
in the ``quark-clustering phase'', where the energy scale could be
high enough to allow the {\em restoration} of light flavor symmetry,
but may not be high enough to make the quarks really deconfined.
Quark-cluster stars are treated here to assort with the type of
quark stars since (1) they manifest in a similar way of quark star
with self-bound rather gravity-bound of neutron stars, (2) their
equation of state should be understood in the level of quarks rather
than hadrons (i.e., the quark degree of freedom would play a
significant role in determining the equation of state and during the
formation of quark-cluster stars), and (3) the term of
``quark-cluster star'' might be abbreviated simply as ``quark
star''.

The observational tests from polarization, pulsar timing and
asteroseismology have been discussed,~\cite{Xu03} and it is found
that the idea of clustering quark matter could provide us a way to
understand different manifestations of pulsars.
The realistic quark stars could then be actually ``quark-cluster
stars''.
An interesting suggestion is that quark matter could be in a solid
state~\cite{Horvath05,Owen05,mrs07}, and for quark-cluster stars,
solidification could be a natural result if the kinetic energy of
quark clusters is lower than the interaction energy between the
clusters. To calculate the interaction between quarks and to predict
the state of matter for quark stars by QCD calculations is a
difficult task; however, it is still meaningful for us to consider
phenomenologically some models to explore the properties of quarks
at the low energy scale. In this paper we show two models for
clustering quark matter. In the Lennard-Jones model~\cite{LX09b} we
take the number of quarks inside each quark-cluster $N_q$ to be a
free parameter, and in the $H$-stars model~\cite{LGX11} we take
$H$-cluster as a specific kind of quark-clusters. Under a wide range
of parameter-space, the maximum mass of quark-cluster stars could be
well above $2M_\odot$.

The asymptotic freedom of QCD makes the perturbative theory
applicable to study the systems under strong interaction, but it
cannot describe the systems with vast assemblies of particles under
strong interaction that exist in the Universe.
Quark matter at high density and low temperature is difficult to be
created in laboratories as well as difficult to be study along with
QCD calculations, and the observational tests should play an
important role to constrain the properties of QCD at low energy
scales.

%\section{From nuclei to quark-cluster stars}

\section{Clustering Quark Matter}

Due to QCD's asymptotic freedom, cold dense quark matter would
certainly be of Fermi gas or liquid if the baryon density is
extremely high. However, perturbative QCD would work reasonably well
only for quark chemical potentials above 1 GeV, while the quark
chemical potential for a typical quark stars is about 0.4 GeV. We
can make an estimate on the chemical potential and the interaction
energy of quarks inside quark stars.

The strong interaction between quarks in compact stars may result in
the formation of quark clusters, with a length scale $l_q$ and an
interaction energy $E_q$. An estimate from Heisenberg's relation, if
quarks are dressed with mass $m_q\simeq 300\ \rm MeV$, gives
$l_q\sim \frac{1}{\alpha}{\hbar c}\ {m_q c^2}\simeq
\frac{1}{\alpha}\ \rm fm$, and $E_q\sim \alpha^2 m c^2\simeq 300
\alpha_s^2\ \rm MeV$. The strong coupling constant $\alpha$ can be
estimated from non-perturbative QCD as a function of baryon number
density, and at a few nuclear density in compact stars, the coupling
could be very strong rather than weak, with $\alpha \simeq
2$~\cite{Xu:2010}. This means that a weakly coupling treatment could
be dangerous for realistic cold quark matter, and quarks would be
clustered.

Quark-clusters could emerge in cold dense matter because of the
strong coupling between quarks.
The quark-clustering phase has high density and the strong
interaction is still dominant, so it is different from the usual
hadron phase, and on the other hand, the quark-clustering phase is
also different from the conventional quark matter phase which is
composed of relativistic and weakly interacting quarks.
The quark-clustering phase could be considered as an intermediate
state between hadron phase and free-quark phase, with deconfined
quarks grouped into quark-clusters, and that is another reason why
we take quark-cluster stars as a special kind of quark stars.
%
%$H$-cluster stars are self-bound due to the interaction between clusters, with non-vanishing surface density but vanishing surface pressure.
%
It is worth noting that, whether the chiral symmetry broken and confinement phase transition happen simultaneously inside compact stars is a matter of debate (see~\cite{Andronic:2009gj} and references therein), but here we assume that the chiral symmetry is broken in quark-clustering phase.

What are quark-clusters explicitly? There is no clear answer, and we could only have some candidates.
A 18-quark cluster, called quark-alpha~\cite{Michel1988}, could be
completely symmetric in spin, light flavor and color space.
$\Lambda$ particles, with structure $uds$, is the light particle with light flavor symmetry.
There could probably attraction between two $\Lambda$s~\cite{Beane:2010hg,Inoue:2010es}, so $H$-clusters with structure $uuddss$ could emerge.
If the light flavor symmetry is ensured, then the dominant components inside the stars is very likely to be $H$-clusters.
In the following, we will show that the degree of light flavor symmetry breaking could be small in the macroscopic strange quark matter.

{\it About light flavor symmetry.} It is well know that there is an asymmetry term to account for the
observed tendency to have equal numbers of protons ($Z$) and
neutrons ($N$) in the liquid drop model of the nucleus. This nuclear
symmetry energy (or the isospin one) represents a symmetry between
proton and neutron in the nucleon degree of freedom, and is actually
that of up and down quarks in the quark degree~\cite{Li_Chen2008}.
The possibility of electrons inside a nucleus is negligible because
its radius is much smaller than the Compton wavelength $\lambda_c =
h/m_ec = 0.24\AA$. The lepton degree of freedom would then be not
significant for nucleus, but electrons are inside a large or gigantic nucleus, which is the case
of compact stars.
Now there is a competition: isospin symmetry favors $Z=N$ while
lepton chemical equilibrium tends to have $Z\ll N$. The nuclear
symmetry energy $\sim 100 (Z-N)^2/A$ MeV, where $A=Z+N$, could be
around 100 MeV per baryon if $N\gg Z$. Interesting, the kinematic
energy of an electron is also $\sim 100$ MeV if the isospin symmetry
keeps in nuclear matter.
However, the situation becomes different if strangeness is included:
no electrons exist if the matter is composed by equal numbers of
light quarks of $u$, $d$, and $s$ with chemical equilibrium.
In this case, the 3-flavor symmetry, an analogy of the symmetry of
$u$ and $d$ in nucleus, may results in a ground state of matter for
gigantic nuclei. Certainly the mass different between $u$, $d$ and $s$ quarks would also break the symmetry, but the interaction between quarks could lower the effect of mass differences and try to restore the symmetry.
Although it is hard for us to calculate how strong the interaction between quarks is, the non-perturbative nature and the energy scale of the system make it reasonable to assume that the degree of the light flavor symmetry breaking is small, and there is a few electrons (with energy $\sim 10$ MeV).

The above argument could be considered as an extension of the
Bodmer-Witten's conjecture. Possibly it doesn't matter whether three
flavors of quarks are free or bound.
We may thus re-define {\em strange matter} as cold dense matter with
light flavor symmetry of three flavors of $u$, $d$, and $s$ quarks.

\section{The Global Structure of Quark-cluster Stars}

We propose that pulsar-like compact stars could be quark-cluster stars which are totally composed of quark clusters.
Quark-cluster stars could have different properties from neutron stars and conventional quark stars, such as the radiation properties, cooling behavior and global structure.
In this paper, we only focus on the global structure of quark-cluster stars, deriving the mass-radius relation based on the equation of state, under the Lennard-Jones model and $H$ star model respectively.

\subsection{Lennard-Jones quark matter model}

In the Lennard-Jones quark matter model, the
interaction potential $u$ between two quark-clusters as the function
of their distance $r$ is assumed to be described by the
Lennard-Jones potential, similar to that between inert molecules,
\begin{equation}u(r)=4U_0[(\frac{r_0}{r})^{12}-(\frac{r_0}{r})^6],\end{equation}
where $U_0$ is the depth of the potential and $r_0$ can be
considered as the range of interaction.
It is worth noting that the property of short-distance repulsion and
long-distance attraction presented by Lennard-Jones potential is
also a characteristic of the interaction between nuclei.

Under the interaction, quark-clusters could be localized and behave like classical particles, and in this case the tatal interaction potential of one quark-cluster could be written as
\begin{equation}U(R)=2U_0[A_{12}(\frac{r_0}{R})^{12}-A_6(\frac{r_0}{R})^6], \label{LJ}\end{equation}
where $R$ is the nearest distance between two quark-clusters, and $A_{12}$ and $A_6$ are the coefficients depending on the lattice structure.
The localization is natural for clustering quark matter, with the following reasons.
One quark-cluster with mass $m$ is under the composition of interaction from its neighbor quark-clusters, which forms a potential well.
The energy fluctuation makes this quark-cluster oscillate about its equilibrium position with the deviation $\Delta x$, $\Delta E\simeq\hbar ^2/(m \Delta x^2)\simeq k \Delta x^2$, where $k\simeq \partial^2V(r)/\partial r^2$, and $r$ is the distance of two neighbor $H$-clusters.
We use the inter-cluster interaction in Eq(\ref{LJ}), and estimate $\Delta x$ at density $\rho=10\rho_0$, $\Delta x \simeq (\hbar^2/m k)^{1/4} \simeq 0.2\ {\rm fm} (18/N_q)^{1/4}$, where $N_q$ is the number of quarks inside each quark-cluster.
%
%The Compton wavelength of an $H$-cluster is $\lambda \sim h/(m_H c)\simeq 0.56\ {\rm fm}\ (2210\ {\rm MeV}/m_H)$.
%
On the other hand, the distance between two nearby quark-clusters at density $\rho=10\rho_0$ is $d=n^{-1/3}\simeq
1.5\ {\rm fm}\ (N_q/18)^{1/3}$, with $n$ the number density of quark-clusters.
Consequently, the interaction would localize $H$-clusters in the potential well at the stellar center, since $\Delta x<d$.
On the stellar surface, $\rho\simeq 2\rho_0$, we have $\Delta x = 0.4$ fm and $d \sim 2.6$ fm.
Therefore, under the interaction, quark-clusters could be localized and behave like classical particles, and the quantum effect would be negligible.

Under such potential, we can get the equation of state for quark-cluster stars.
Because of the strong interaction, the surface density $\rho_s$
should be non-zero, and $r_0$ can be derived at the surface where the pressure vanishes.
Choosing $\rho_s=2\rho_0$ to ensure the deconfinement of quarks, we can derive the mass and radius of a quark star by combining the equation of state with the hydrostatic equilibrium condition.
The dependence of maximum mass of quark stars on  $U_0$ and $N_q$ is shown in Figure 2~\cite{LX11a}.
We can see that there is enough parameter space
for the existence of quark stars with mass larger than $2M_\odot$.
The case $N_q>10^4$ should be ruled out by the discovery of PSR J1614-2230.

%%%%%%%%%%%%%%%%%%%%%%%%%%%%%%%%%%%%%%%%%%%%%%%%%%%
\begin{figure}[th]
\centerline{\psfig{file=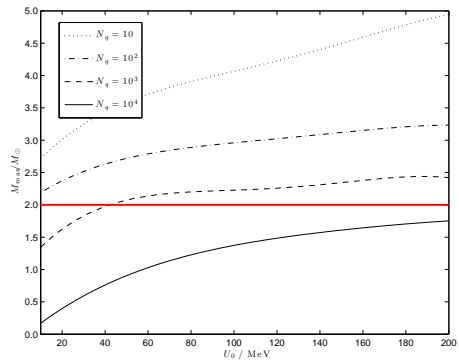,width=7cm}}
\vspace*{8pt}
\caption{The dependence of maximum mass $M_{\rm max}$ on $U_0$ (depth of
potential well), for some different cases of $N_q$ (number of quarks
inside one quark-cluster), in Lennard-Jones cold quark matter model.
The surface density $\rho_s$ is chosen to be 2 times of $\rho_0$
(the nuclear matter density).}
\end{figure}
%%%%%%%%%%%%%%%%%%%%%%%%%%%%%%%%%%%%%%%%%%%%%%%%%%%

\subsection{$H$-cluster Stars}

In Lennard-Jones model, quark-clusters are analogized to electric neutral molecules; however, quark clusters may also be analogized to hadrons.
A dihyperon with quantum numbers of $\Lambda\Lambda$ ($H$ dibaryon)
was predicted to be stable state or resonance~\cite{Jaffe1977}, and recent lattice QCD simulations show some evidence that the $H$-dibaryons (with structure $uuddss$) are bound states~\cite{Beane:2010hg,Inoue:2010es}.
Motivated by both the theoretical prediction and numerical simulations, we consider a possible kind
of quark-clusters, $H$-particles, that could emerge inside quark
stars during their cooling, as the dominant building blocks~\cite{LGX11}.
To study quark stars composed of $H$-matter, i.e. $H$ stars, we assume that the interaction between $H$-particles is mediated by $\sigma$ and $\omega$ mesons and introduce the Yukawa potential to describe the $H$-$H$ interaction~\cite{Faessler1997}, and then derive the dependence of the maximum mass of $H$ stars on the depth of potential well, taking into account the in-medium stiffening effect.

Using the similar argument as that in the Lennard-Jones model, $H$-clusters could be localized and behave like classical particles, and Bose condensate would not take place even though they are bosons.
On the other hand, although $H$-clusters could be weakly bound particles which would decay to lighter baryons, such as the reaction of $H\rightarrow 2n+2\pi$, the decay could hardly happen inside compact stars.
At density $\rho$ larger than $2\rho_0$, the fermi energy of neutrons is larger than 100 MeV, which makes $H$-clusters difficult to decay into neutrons and pions.
Moreover, $H$-clusters could be safe under the high momentum fluctuation $\Delta p$ at high densities inside compact stars, because the energy fluctuation $\Delta E$ is not so high due to their high mass.
We can make the estimation of $\Delta E\sim \Delta p^2/2m_H\simeq 7\ {\rm MeV} (\rho/10\rho_0)^{2/3}(m_H/2210\ \rm MeV)^{-5/3}$, where we set the mass of $H$-cluster, $m_H=2m_\Lambda-20\ \rm
MeV=2210\ \rm MeV$, and $m_\Lambda$ the mass of $\Lambda^0$.
The energy of $\Delta E$ could be not much lower than the binding energy of $H$-clusters and potential drop of interaction between $H$-clusters, but it could be reasonable to ensure the existence of $H$-clusters with large enough mass fraction of the star.

Compact stars composed of pure $H$-clusters are electric neutral, but in reality there could be some flavor symmetry breaking that leads to the non-equality among $u$, $d$ and $s$, usually with less $s$ than $u$ and $d$.
The positively charged quark matter is necessary because it allows the existence of electrons that is crucial for us to understand the radiative properties of pulsars.
The pressure of degenerate electrons is negligible compared to the pressure of $H$-clusters, so the contribution of electrons to the equation of state is negligible.

We adopt the Yukawa potential with $\sigma$ and $\omega$ coupling between $H$-particles~\cite{Faessler1997},
\begin{equation} V(r)=\frac{g_{\omega H}^2}{4\pi}\frac{e^{-m_\omega^* r}}{r}-\frac{g_{\sigma H}^2}{4\pi}\frac{e^{-m_\sigma^* r}}{r},\label{V}
\end{equation}
where $g_{\omega H}$ and $g_{\sigma H}$ are the coupling constants of $H$-particles and meson fields.
In dense nuclear matter, the in-medium stiffening effect, i.e., the Brown-Rho scaling effect, should be considered~\cite{BR2004}, and then the effective meson masses $m_M^*$ satisfy the scaling law $m_M^*\simeq m_M(1-\alpha_{BR} n/n_0)$, where $\alpha_{BR}$ is the coefficient of the scaling and $m_M$ is the meson mass in free space.
In the problem we are now considering, however, a quark star is at supra-nuclear density, and we then use a modified scaling law of
\begin{equation} m_M^*=m_M \exp(-\alpha_{BR}n/n_0),
\end{equation}
which also shows the in-medium effect that stiffens the inter-particle potential by reducing the meson effective masses, and approximately the same as the usual scaling law at the nuclear matter density.

From the potential, we can get the equation of state, and derive the total mass $M$ and radius $R$ of an $H$ star by numerical integration.
%Figure. 3 shows the mass-radius and mass-central density (rest-mass energy density) curves, in the case $\rho_s=2 \rho_0$ and $\alpha_{BR}$=0.2, including $V_0=-10$ MeV (solid line) and $V_0=-100$ MeV (dashed line).
Figure 3~\cite{LGX11} shows the dependence of $M_{\rm max}$ on the depth of the potential well $V_0$ and the Brown-Rho scaling coefficient $\alpha_{BR}$, in the case $\rho_s=2\rho_0$.
To make comparison, we also plot the result when $\alpha_{BR}=0$, and the discrepancy between different values of non-zero $\alpha_{BR}$ is not very significant.

%%%%%%%%%%%%%%%%%%%%%%%%%%%%%%%%%%%%%%%%%%%%%%%%%%%
\begin{figure}[th]
\centerline{\psfig{file=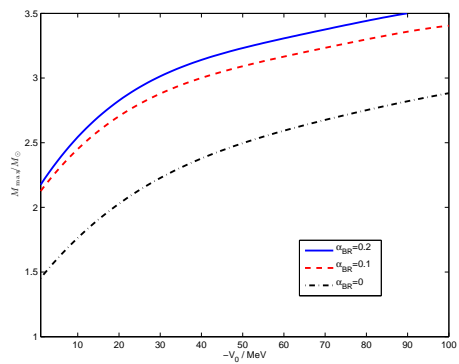,width=7cm}}
\vspace*{8pt}
\caption{The dependence of $M_{\rm max}$ on $V_0$ and $\alpha_{BR}$, in the case $\rho_s=2\rho_0$, including $\alpha_{BR}=0.5$ (solid line), $\alpha_{BR}=0.2$ (dashed line) and $\alpha_{BR}=0$ (dotted line).}
\end{figure}
%%%%%%%%%%%%%%%%%%%%%%%%%%%%%%%%%%%%%%%%%%%%%%%%%%%

$H$ stars could have stiff equation of state, and under a wide range of parameter-space, the maximum mass of $H$ stars can be well above 2$M_\odot$, providing a possible way to explain the observed high mass of the newly discovered pulsar PSR J1614-2230.
Furthermore, if we know about the properties of pulsars from observations, we can
get information on $H$-$H$ interaction; for example, if a pulsar
with mass larger than $3M_\odot$ is discovered, then we can
constrain $-V_0$ to be larger than 60 MeV.

\section{Conclusions and Discussions}

Pulsars could be either neutron stars or quark stars.%
~\footnote{%
Strictly speaking, quark-cluster stars, where the degree of freedom
is quark cluster, are {\em not} traditional quark stars if one
thinks that the latter are composed by free quarks. Nonetheless, in
this paper, we temporarily consider quark-cluster stars as a very
special kind of quark stars. In astrophysics, it is evident that
quark-cluster stars manifest themselves similar to quark stars
rather than neutron stars.
} %
Although the state of cold quark matter at a few nuclear densities
is still an unsolved problem in low energy QCD, various pulsar
phenomena would give us some hints about the properties of elemental
strong interaction, complementary to the terrestrial experiments.
Pulsar-like compact stars provide high density and relatively low
temperature conditions where quark matter with quark-clusters could
exist, and we have discussed some possible kinds of models to
describe such kind of quark matter which could be tested by
observations. We apply the Lennard-Jones model and $H$ star model,
where the quark-clusters are treated as molecules in the former and
dibaryons in the latter. The $2M_\odot$ pulsar puts constraints on
the number of quarks in one quark-cluster $N_q$ to be less than
$10^4$ in the Lennard-Jones model. To put any constraint on the
$H$-matter model with in-medium stiffening effect, some more massive
pulsars (e.g. $M>3M_\odot$) should be found in the future.

After addressing a lot about modeling quark-cluster stars, it could
be interesting to compare Landau's ``giant nucleus'', neutron star
and quark-cluster star. Landau conjectured a ``neutronic'' state
core with nuclear matter densities inside a star in order to solve
the origin problem of stellar energy. In Landau's scenario, the
``neutronic'' state core and the surrounding ordinary matter are in
chemical equilibrium at the boundary, which is very similar to the
neutron star picture where the inner and outer cores and the crust
keep chemical equilibrium at each boundary. Landau's giant nucleus
is then bound by gravity. The ``neutronic'' core should have a
boundary and is in equilibrium with the ordinary matter because the
star has a surface composed of ordinary matter. There is, however,
no clear observational evidence for a neutron star's surface,
although most of authors still take it for granted that there should
be ordinary matter on surface, and consequently a neutron star has
different components from inner to outer parts. Being similar to
traditional quark stars, quark-cluster stars have almost the same
composition from the center to the surface, and the quark matter
surface could be natural for understanding some different
observations. As an analog of neutrons, quark-clusters are bound
states of several quarks, so to this point of view a quark-cluster
star is more similar to a {\em real} giant nucleus of self-bound
(not that of Landau), rather than ``giant hadron'' which describes
traditional quark stars.

It is also worth noting that, although composed of quark-clusters,
quark-cluster stars are self-bound. They are bound by the residual
interaction between quark-clusters. This is different from but
similar to the traditional MIT bag scenario. The interaction between
quark-clusters could be strong enough to bind the star, and on the
surface, the quark-clusters are just in the potential well of the
interaction, leading to non-vanishing density but vanishing
pressure.

It has been 80 years since Landau proposed the idea of ``neutron''
stars, and more than 40 years since the first pulsar was discovered,
but the interior structure of pulsar-like compact stars is still in
controversy. The nature of pulsars is essentially a non-perturbative
QCD problem, corresponding the region between hadron phase and
quark-gluon plasma phase in the QCD phase diagram. Although the
state of cold quark matter at a few nuclear densities is still an
unsolved problem in low energy QCD, various pulsar phenomena would
give us some hints about the properties of elemental strong
interaction,~\cite{Xu:2010} complementary to the terrestrial
experiments. Pulsar-like compact stars provide high density and
relatively low temperature conditions where quarks may not be free
but would be clustered to form quark-cluster matter.
Whether this quark matter composed of quark-clusters could achieve
at supra-nuclear density is still unknown, and on the other hand,
the nature of pulsar-like stars also depends on the physics of
condensed matter. These problems are essentially related to the
non-perturbative QCD, and we hope that future astrophysical
observations would test the existence of quark-cluster stars.

\section*{Acknowledgements}

We would like to thank useful discussions at our pulsar group of
PKU. This work is supported by the National Natural Science
Foundation of China (Grant Nos. 10935001, 10973002), the National
Basic Research Program of China (Grant Nos. 2009CB824800,
2012CB821800), the John Templeton Foundation, and China Postdoctoral
Science Foundation Project.

%\begin{thebibliography}{000} %for 3 digits
%\begin{thebibliography}{00}  %for 2 digits

\end{document}